%% file: proceedings.tex
\documentclass[a4paper,11pt]{article}
\usepackage{pos}

\input{definitions.tex}

\title{Padé and Padé-Laplace Methods for masses and matrix elements}

\author*[a]{Georg M. von Hippel}

\affiliation[a]{PRISMA$^+$ Cluster of Excellence and Institute of Nuclear
Physics, University of Mainz,\\
  Johann-Joachim-Becher-Weg 45, 55099 Mainz, Germany}

\emailAdd{hippel@uni-mainz.de}

\abstract{The problem of having to reconstruct the decay rates and
corresponding amplitudes of the single-exponential components of a noisy
multi-exponential signal is common in many other areas of physics and
engineering besides lattice field theory, and it can be helpful to study the
methods devised and used for that purpose in those contexts in order to get a
better handle on the problem of extracting masses and matrix elements from
lattice correlators. Here we consider the use of Padé and Padé-Laplace methods,
which have found wide use in laser fluorescence spectroscopy and beyond,
emphasizing the importance of using robust Padé approximants to avoid spurious
poles. To facilitate the accurate evaluation of the Laplace transform required
for the Padé-Laplace method, we also present a novel approach to the numerical
quadrature of multi-exponential functions.
}

\FullConference{%
  The 39th International Symposium on Lattice Field Theory (Lattice2022),\\
  8-13 August, 2022 \\
  Bonn, Germany 
}

\begin{document}
\maketitle

\section{Introduction}

A crucial problem in lattice field theory is the extraction of particle masses
$E_n$ and matrix elements $\braket{n}{\hat{O}}{0}$ from measured correlation
functions via their spectral representation,
\begin{equation}
C(t) = \langle O(t)O^*(0)\rangle \nonumber
= \sum_{n=1}^\infty A_n\rme^{-E_n t}
\label{eq:spectral}
\end{equation}
with $A_n=\left|\braket{n}{\hat{O}}{0}\right|^2$.

While this can in principle be achieved by fitting the measured data,
multiexponential fits tend to be unstable and typically require stabilization
with Bayesian priors. While variational methods work well for a prior-free
determination of masses and matrix elements, the require a full correlator
matrix with a range of operators $O_i$ at both source and sink, which is not
always a feasible option.

It is therefore desirable to have at one's disposal methods that can solve this
problem with just one operator at source and sink. To this end, it is useful to
consider methods used for similar purposes in other fields, such as laser
fluorescence spectroscopy \cite{Bajzer:1989}. Previously, Prony's method found
its way into the lattice toolkit in a similar manner from NMR spectroscopy
\cite{Fleming:2004hs}.

\section{The Padé Method}

The correlator $C(t)$ is measured only on discrete timeslices, $C(ka)$,
$k\in\Nset$. A natural object to study is thus the $Z$ transform of
$C=\{C(ka)~|~k\in\Nset\}$ given by
\begin{equation}
\mathcal{Z}[C](z) = \sum_{k=0}^\infty C(ka) z^{-k} 
=\sum_{n=1}^\infty \frac{A_n z}{z-\lambda_n}
\end{equation}
with $\lambda_n = \rme^{-E_n a}$. One immediately notes that the poles and
residues of $\mathcal{Z}[C]$ give the masses and matrix elements.

However, in practice, only finitely many values $C(ka)$ of the correlator are
known, which renders the series defining $\mathcal{Z}[C]$ inaccessible.
Nevertheless, one can consider Padé approximants to $\mathcal{Z}[C]$ and use
their poles and residues as estimates of $\lambda_n$, $A_n$. Interestingly, a
naive implementation of this idea is equivalent to Prony's method in exact
arithmetic \cite{Weiss:1963}.

However, the results tend to be very unstable in the presence of even minimal
round-off error: common issues include the emergence of pairs of
complex-conjugate poles off the real axis, and of Froissart doublets (a pole
and a nearby zero which fail to cancel each other, typically signalled by
anomalously small residuals). A solution to these problems is given by the
method of robust Padé approximants \cite{Gonnet:2013}, which use the singular-value
decompositions to give an estimate of how many states can be identified from
the data.

\begin{figure}
\begin{center}
\includegraphics[width=0.67\linewidth,keepaspectratio=]{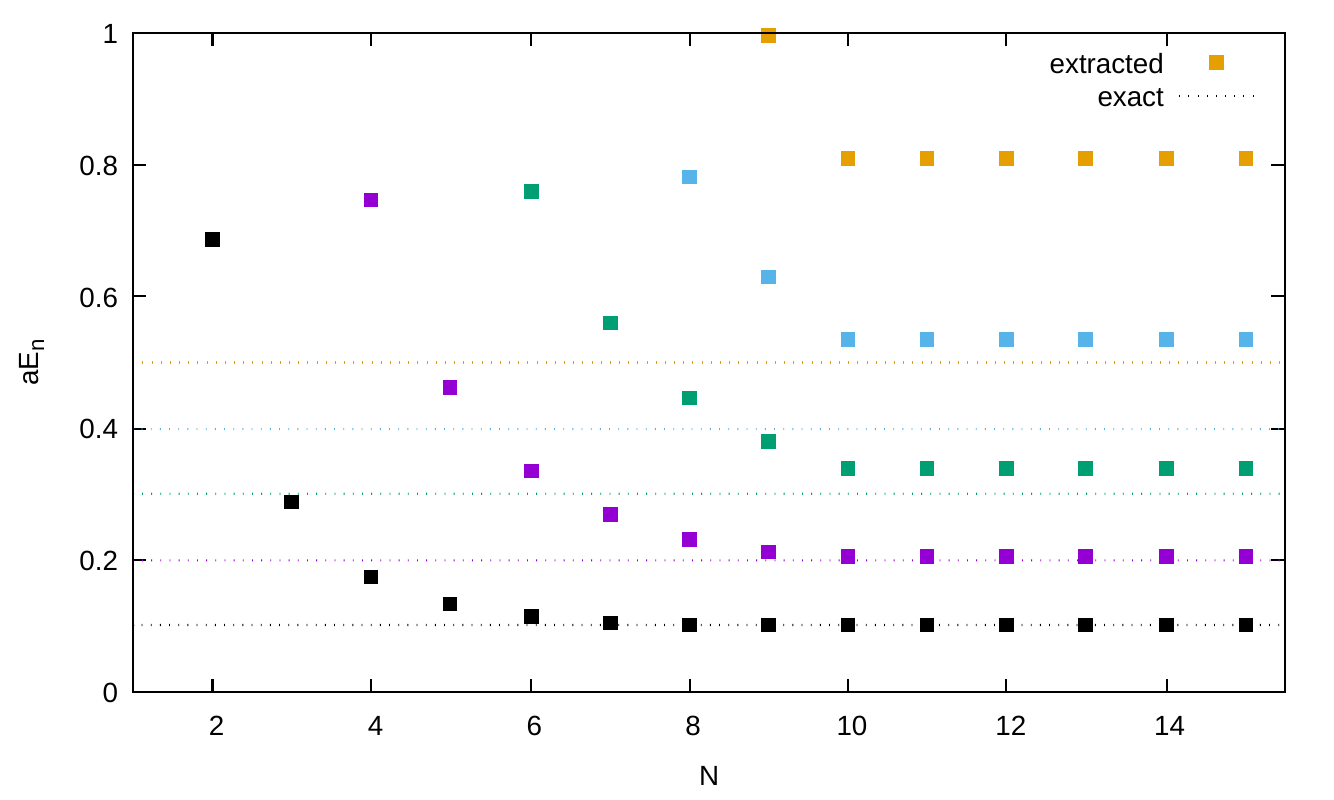}
\end{center}
\caption{\label{fig:pade1} Results from the Padé Method with robust approximants
   applied to synthetic data. Dashed horizontal lines denote the exact values
   of the masses, points the extracted results as a function of the number $N$
   of poles tried.}
\end{figure}

\begin{figure}
\begin{center}
\includegraphics[width=0.67\linewidth,keepaspectratio=]{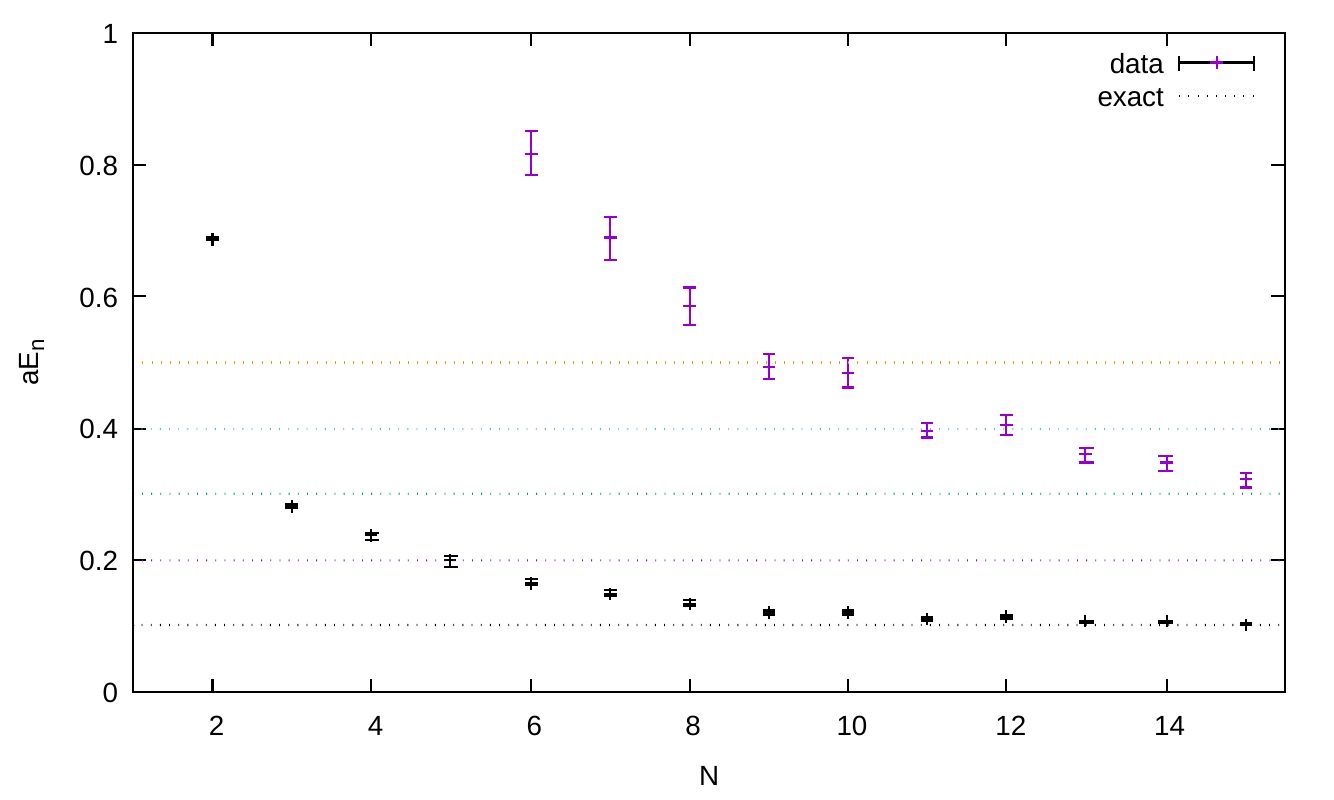}
\end{center}
\caption{\label{fig:pade2} Same as Figure \ref{fig:pade1}, but with a 1\% noise
   applied to the synthetic data.}
\end{figure}

Even with robust methods, the results obtained using the Padé method are known
to quickly deteriorate in the presence of noise
\cite{Gilewicz:1997,Gilewicz:199}. This is borne out by
numerical tests on synthetic data. In Fig.~\ref{fig:pade1}, we plot the
reconstructed energy levels on a synthetic data set with equally spaced energy
levels as a function of the number of poles used for the Padé approximant; it
can be clearly seen that the lowest few energy levels exponentially approach
their true values as the number of poles is increased, until the robust
procedure eventually saturates the number of extracted poles. In contrast,
Fig.~\ref{fig:pade2} shows the results of applying the same procedure to the same
toy model, but with a 1\% noise applied to the synthetic data; here, the
exponential approach to the true value is much slower, and only one excited
state can be extracted, albeit with a large systematic error on its energy.

\section{The Padé-Laplace Method}

To avoid the shortcomings of the Padé method, the Padé-Laplace method was
proposed by Yeramian and Claverie in \cite{Yeramian:1987}. It is based on noting that the
Laplace transform of the correlation function $C(t)$ is given by
\begin{equation}
\mathcal{L}[C](p) = \int_0^\infty \rmd t\,C(t)\rme^{-pt}
= -\sum_{n=1}^\infty \frac{A_n}{p+E_n}
\end{equation}
so that the poles and residues of the Laplace transform again yield the masses
and matrix elements of interest.

The full functional form of the Laplace transform is not available from
numerical data, but we can compute Padé approximants to $\mathcal{L}[C]$
by noting that the Taylor coefficients of $\mathcal{L}[C]$ are just the moments
of $C(t)$,
\begin{align}
\left.\frac{\rmd^k}{\rmd p^k}\mathcal{L}[C](p)\right|_{p=p_0} = \int_0^\infty
\rmd t\,(-t)^k C(t)\rme^{-p_0t}
\label{eq:moments}
\end{align}
and use their poles and residues as estimates for $E_n$, $A_n$.

Since the integrals for the moments cannot be analytically evaluated given the
numerical data $C(ka)$, the moments need to be estimated using quadrature
formulae. As a result, robust methods are even more important in the case of
the Padé-Laplace method due to the numerical errors arising from numerical
integration.

\begin{figure}
\begin{center}
\includegraphics[width=0.67\linewidth,keepaspectratio=]{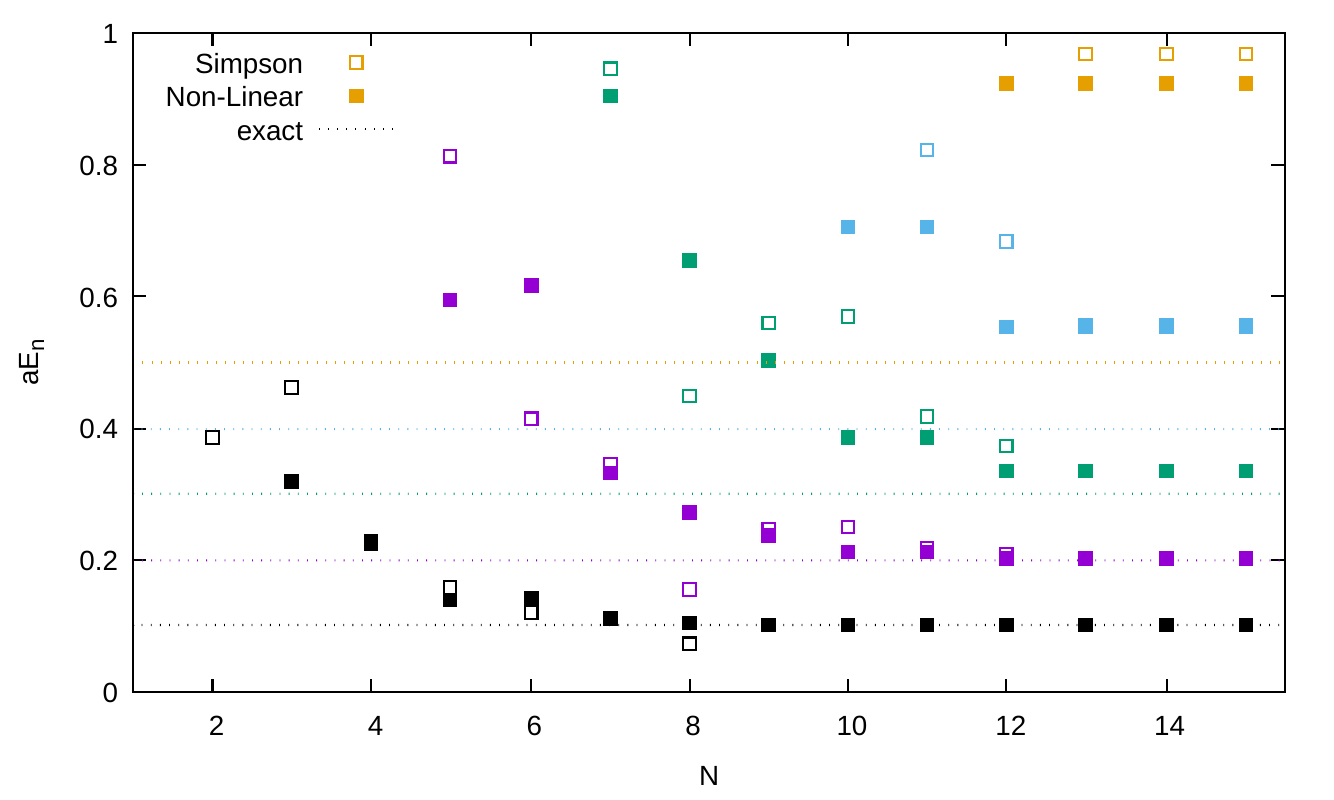}
\end{center}
\caption{\label{fig:padelap1} Results from the Padé-Laplace Method with robust
   approximants applied to synthetic data. Dashed horizontal lines denote the
   exact values of the masses, points the extracted results as a function of
   the number $N$ of poles tried. Open symbols use Simpson's rule for the numerical
   integration, filled symbols a novel non-linear quadrature formula specialized
   to exponentially-decaying integrands.}
\end{figure}

\begin{figure}
\begin{center}
\includegraphics[width=0.67\linewidth,keepaspectratio=]{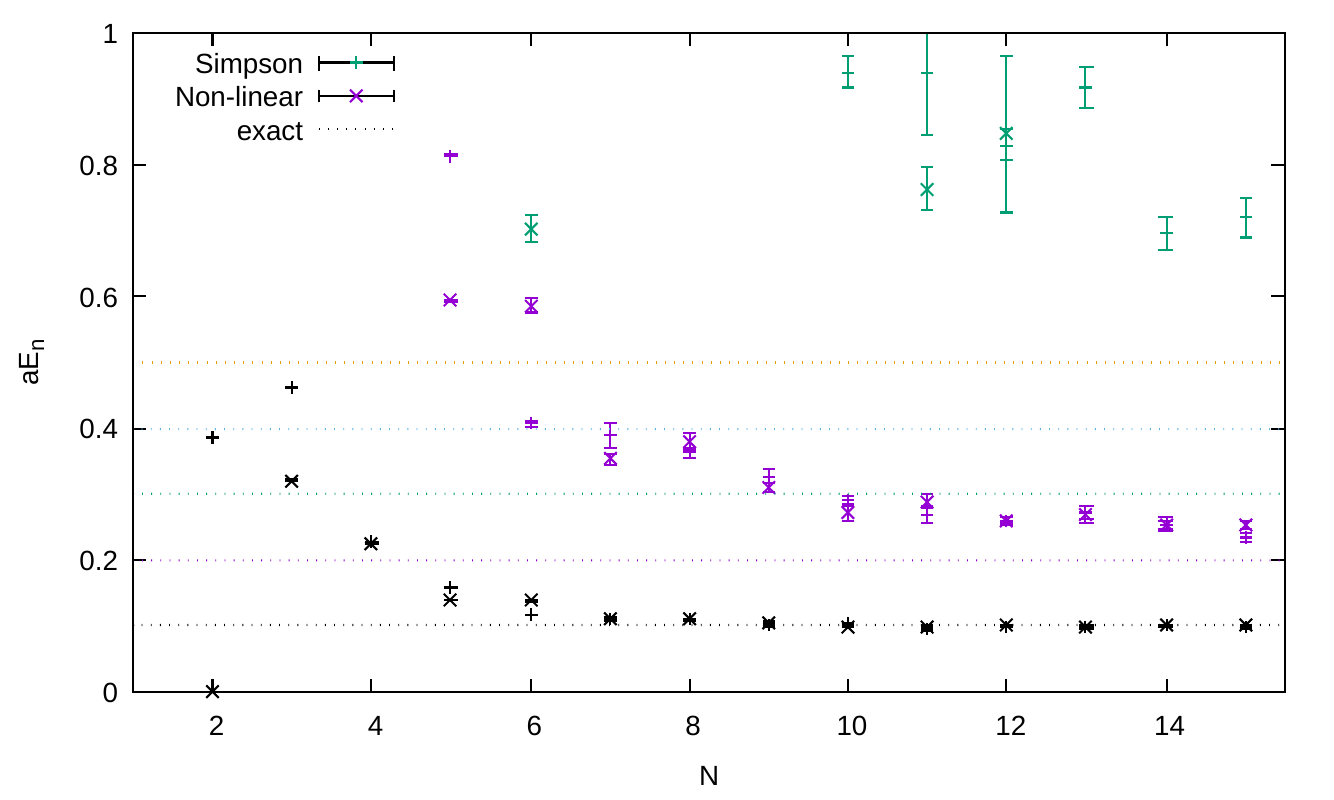}
\end{center}
\caption{\label{fig:padelap2} Same as Figure \ref{fig:padelap1},
   but with a 1\% noise applied to the synthetic data.}
\end{figure}

Due to the averaging over different times that is inherent in the Laplace
transform, the results from the Padé-Laplace method tend to be more resistant
to noise than those from the Padé method. As an example,
Fig.~\ref{fig:padelap1} shows the results of applying the Padé-Laplace method
to the same synthetic data as in Fig.~\ref{fig:pade1}; a similar approach to
the true value can be observed as the number of poles tried is raised, although
the approach is less monotonic, largely due to the effects of numerical error
from the quadrature formula used, as can be seen from the difference between
the results using Simpson's rule (shown as open symbols) and those using a
novel non-linear quadrature formula of the kind discussed in the next section
(shown as filled symbols). On the other hand, Fig.~\ref{fig:padelap2} shows
the results of applying the Padé-Laplace method to the same noisy synthetic
data as in Fig.~\ref{fig:pade2}; here, the approach to the true value is faster
than for the Padé method, and the first excited state can be extracted much more
reliably (while higher excited states are still drowned out by the noise).

\section{Non-Linear Quadrature Formulae}

When computing moments (\ref{eq:moments}) numerically, e.g. using
the trapezoidal rule or Simpson's rule, the numerical error from quadrature
introduces an additional source of instability to the Padé approximants. To
avoid this source of error as far as possible, it is useful to consider
quadrature formulae that give exact results when applied to functions of the
form (\ref{eq:spectral}).

Unfortunately, the traditional quadrature formulae all are exact by design on
linear spaces of polynomials, while the functions we are interested in
integrating here form a non-linear family. We therefore need to go beyond
traditional quadrature formulae and consider non-linear approximations
\begin{equation}
\int_a^b \rmd x\,f(x) \approx (b-a)\,q(f(a),f(b))
\label{eq:nlq}
\end{equation}
to definite integrals.
The core result \cite{vonHippel:2022qme}
in this context is that as long as (\ref{eq:nlq}) is exact on
all multiples of at least one function, its
integration error is no worse than that of the trapezoidal rule
(and better for functions close to such multiples).

Specifically, a non-linear quadrature formula that is
exact on functions of the form $A\rme^{-m x}$ is given by
\begin{equation}
q(f(a),f(b)) = \frac{f(a)-f(b)}{\log\frac{f(a)}{f(b)}}
\label{eq:nlqexp}
\end{equation}
and one can systematically develop higher-order non-linear quadrature rules
using multiple nodes. For further details and proofs, the reader is referred to
Ref.~\cite{vonHippel:2022qme}.

\section{Conclusions and Outlook}

The Padé-Laplace method has seen wide usage in other
areas of science ranging from biophysics
\cite{Bajzer:1989} and medical research \cite{Scamp:1990} to
climate science \cite{Enting:2022} and food science
\cite{Lodi:2007}. Its application to lattice QCD data is therefore warranted.
It remains to be seen how helpful the newly developed non-linear quadrature
rules will be in this context.

\section*{Acknowledgements}
The author acknowledges useful discussions with Jiyoung Kim and George Fleming.
The author's travel to this conference was partially funded by DFG grant
HI~2048/1-2 (Project No.\ 399400745).

\end{document}

%% file: definitions.tex
\newcommand{\rme}{\textrm{e}}
\newcommand{\rmd}{\textrm{d}}

\newcommand{\Nset}{\mathbb{N}}

\def\sla#1{\setbox0=\hbox{$#1$}                 
   \dimen0=\wd0                                 
   \setbox1=\hbox{/} \dimen1=\wd1               
   \ifdim\dimen0>\dimen1                        
      \rlap{\hbox to \dimen0{\hfil/\hfil}}      
      #1                                        
   \else                                        
      \rlap{\hbox to \dimen1{\hfil$#1$\hfil}}   
      /                                         
   \fi}                                         %
\def\openone{\hbox{\upshape \small1\kern-3.3pt\normalsize1}}

\newcommand{\braket}[3]{\langle#1|#2|#3\rangle}

